\documentclass[dvipdfmx]{PoS}

\title{L\"uscher's finite volume test for two-baryon systems with attractive interactions}

\ShortTitle{L\"uscher's finite volume test for two-baryon systems with attractive interactions}

\author{\speaker{Sinya Aoki} \\
      Center for Gravitational Physics, Yukawa Institute for Theoretical Physics, Kyoto University, Kitashirakawa Oiwakecho, Sakyo-ku, Kyoto 606-8502, Japan, and\\
      Center for Computational Sciences, University of Tsukuba, Tsukuba 305-8577, Japan\\
      E-mail: \email{saoki@yukawa.kyoto-u.ac.jp}}

\author{Takumi Doi\\
Theoretical Research Division, Nishina Center, RIKEN, Wako 351-0198, Japan\\
E-mail: \email{doi@ribf.riken.jp}
}

\author{Takumi Iritani\\
Department of Physics and Astronomy, Stony Brook University, NY 11794-3800, USA, and \\
Theoretical Research Division, Nishina Center, RIKEN, Wako 351-0198, Japan\\
E-mail: \email{takumi.iritani@stonybrook.edu}
}
\author{
for  HAL QCD Collaboration
\begin{center}
\includegraphics[width=0.35\textwidth]{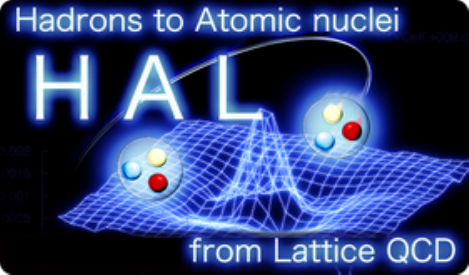}
\end{center}
}

\abstract{
For the attractive interaction,   
the L\"uscher's finite volume formula gives the phase shift at negative squared moment $k^2<0$ for the ground state in the finite volume, which corresponds to the analytic continuation of the phase shift at $k^2<0$ in the infinite volume.
Using this fact, we reexamine behaviors of phase shifts at $k^2 <0$ obtained directly from plateaux of effective energy shifts in previous lattice studies for two nucleon systems on various volumes.
We have found that data, based on which existences of the bound states are claimed, show singular behaviors of the phase shift at $k^2<0$,  which seem incompatible with smooth behaviors predicted by the effective range expansion.
This, together with the fake plateau problem for the determination of the energy shift,
 brings a serious doubt on existences of the $NN$ bound states claimed in previous lattice studies
at  pion masses heavier than 300 MeV.
 }

\FullConference{34th annual International Symposium on Lattice Field Theory\\
		 24-30 July 2016\\
		 University of Southampton, UK}

\begin{document}

\section{Introduction}
Hadron interactions have been investigated by two methods in lattice QCD.
One is the standard direct method, which employs the L\"uscher's  formula\cite{Luscher:1990ux} to calculate the scattering phase shift from the energy of two hadrons in the finite box.
The other is the potential method proposed in Refs.~\cite{Ishii:2006ec,Aoki:2008hh,Aoki:2009ji} and actively applied to various hadron interactions by the HAL QCD collaboration\cite{Aoki:2011ep,Aoki:2012tk}.

The two methods are theoretically  equivalent and  indeed give consistent results for the $I=2$ $\pi\pi$ system\cite{Kurth:2013tua}. On the other hand, the two methods give different conclusions on the nature of the $NN$ interactions at heavier pion masses, as summarized in Fig. 8 in Ref.\cite{Doi:2012ab}. 
While the direct method indicates the existence of the $NN$ bound state in both ${}^1S_0$ (dineutron) and ${}^3S_1$ (deuteron) channels\cite{Yamazaki:2011nd,Yamazaki:2012hi,Yamazaki:2015asa,Beane:2011iw,Beane:2012vq, Orginos:2015aya,Berkowitz:2015eaa}, the HAL QCD method predicts its absence in both channels\cite{Ishii:2006ec,Aoki:2008hh,Aoki:2009ji,Aoki:2011ep,Aoki:2012tk,HALQCD:2012aa}.

\begin{figure}[tbh]
\centering
  \includegraphics[width=0.45\textwidth]{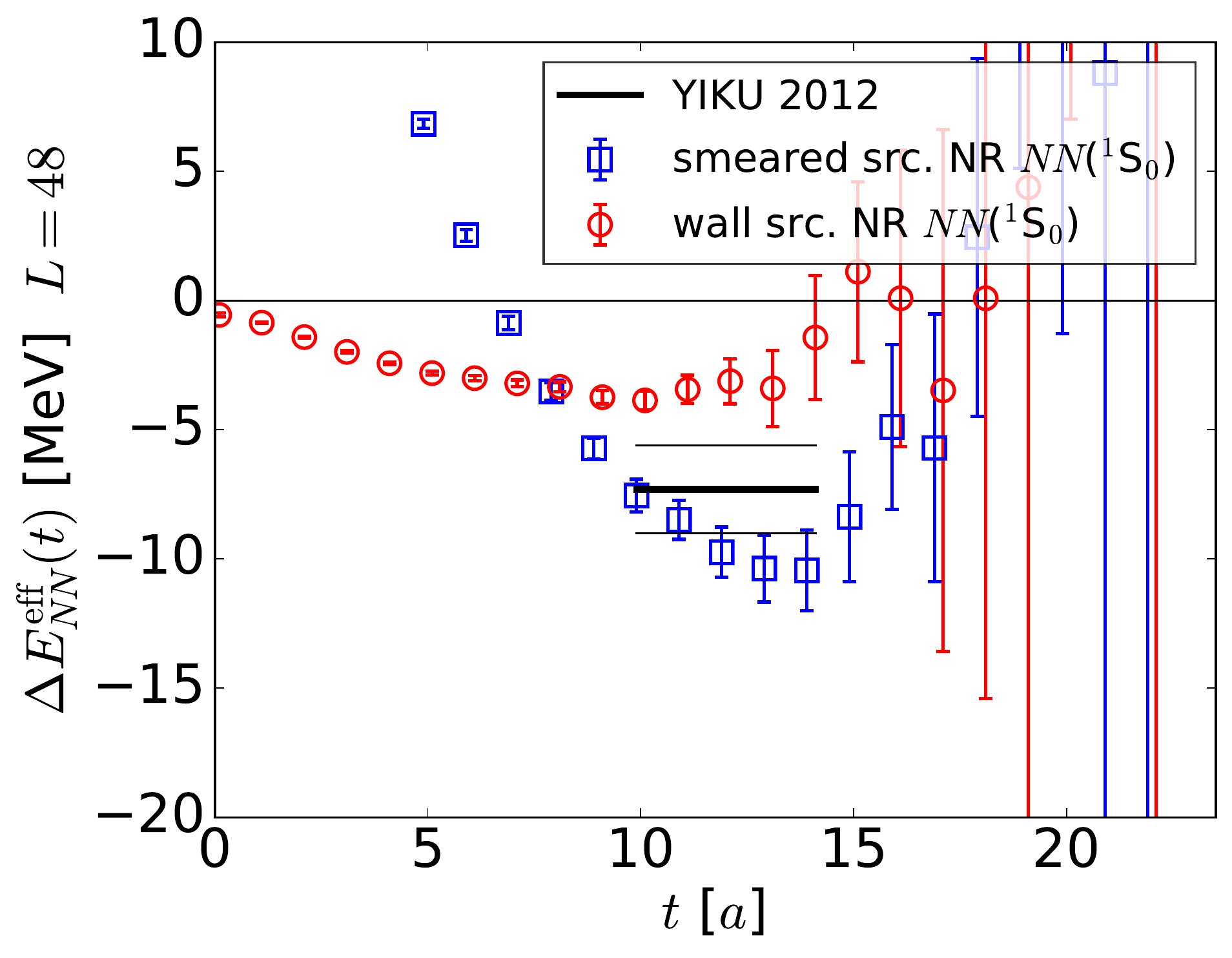}
  \includegraphics[width=0.45\textwidth]{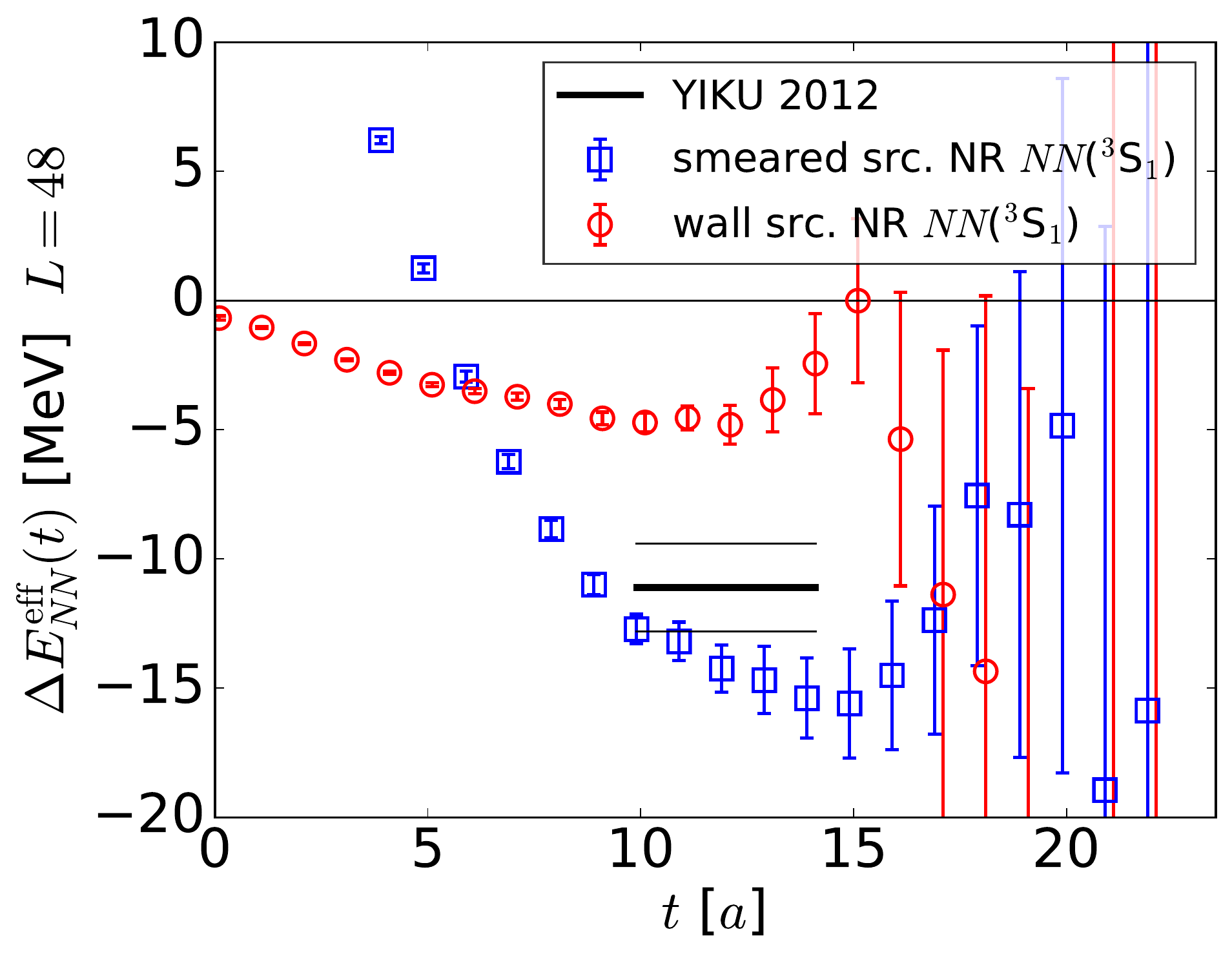} 
 \caption{ Effective energy shift, $\Delta E_{NN}^{\rm eff}(t) = E_{NN}^{\rm  eff}(t) -2 m_N^{\rm eff}(t)$, in the $NN$ (${}^1S_0$) channel (Left) and the $NN$(${}^3S_1$) channel (Right)  for the smeared source (blue squares) and the wall source (red circles). The black solid line represents the fit to the plateau of data in Ref.~\cite{Yamazaki:2012hi} using the same smeared source, where same gauge configurations have been used  to calculate $\Delta E_{NN}^{\rm eff}(t)$.
  }
 \label{fig:E_eff}
\end{figure}

The problem of the direct method has been recently pointed out\cite{Iritani:2015dhu,Iritani:2016jie,Iritani:2016}  that the contamination from excited states at 10\% level can produce the plateau structure of the effective energy shift 
of two baryons, which deviates from the correct plateau. Furthermore,  two different plateau  structures are indeed observed in lattice QCD data, as shown in Fig.~\ref{fig:E_eff}, where the plateaux from smeared source and wall source are compared.

Since the above problem brings a serious doubt on results obtained from the direct method in the literatures, we propose another simpler test for  the energy shift  of two baryons using the L\"uscher's finite volume formula in this report.

\section{Method}
One can extract the scattering phase shift from the 2 particle energy in the finite box using the L\"uscher's formula\cite{Luscher:1990ux}. In the case of the S-wave scattering of two baryons with mass $m_B$ in the center of mass frame,
the phase shift $\delta_0(k)$ is given by
\begin{equation}
k \cot \delta_0 (k) = \frac{1}{\pi L} \sum_{\vec n\in \mathbf{Z}^3}\frac{1}{\vec n^2 -q^2},
\qquad q=\frac{k L}{2\pi}, \quad \Delta E = 2\sqrt{k^2+m_B^2} - 2m_B 
\label{eq:kcot_delta}
\end{equation}
where 
$\Delta E= E_{BB} - 2m_B$ and $E_{BB}$ is the energy of the two baryon state measured in lattice QCD on a finite box with the spatial extension $L$.  
The result provides us informations about the interactions 
through the effective range expansion (ERE) as
\begin{equation}
k\cot\delta_0 (k) = \frac{1}{a_0} + \frac{r_0}{2} k^2+ \sum_{n=2}^\infty b_n k^{2n},
\label{eq:ERE}
\end{equation}
where $a_0$ is the scattering length and $r_0$ is the effective range. 
Since $\Delta E < 0$ for  the attractive interaction, the L\"uscher's formula gives the analytic continuation of $k\cot\delta_0(k)$ at unphysical $k^2 <0$, which should also follow the ERE curve determined at physical $k^2>0$.

\begin{figure}[tbh]
\centering
  \includegraphics[width=0.49\textwidth]{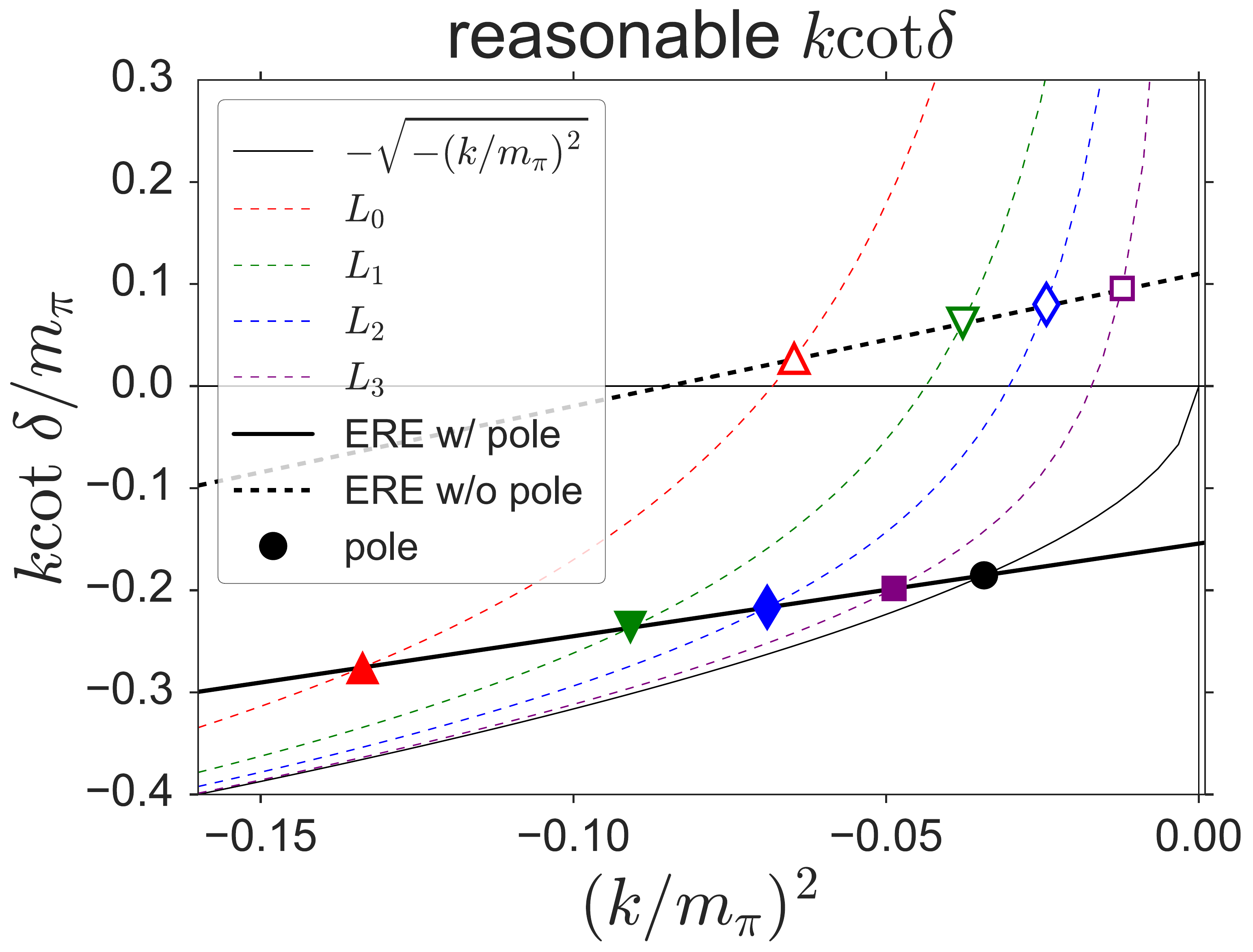}
   \includegraphics[width=0.49\textwidth]{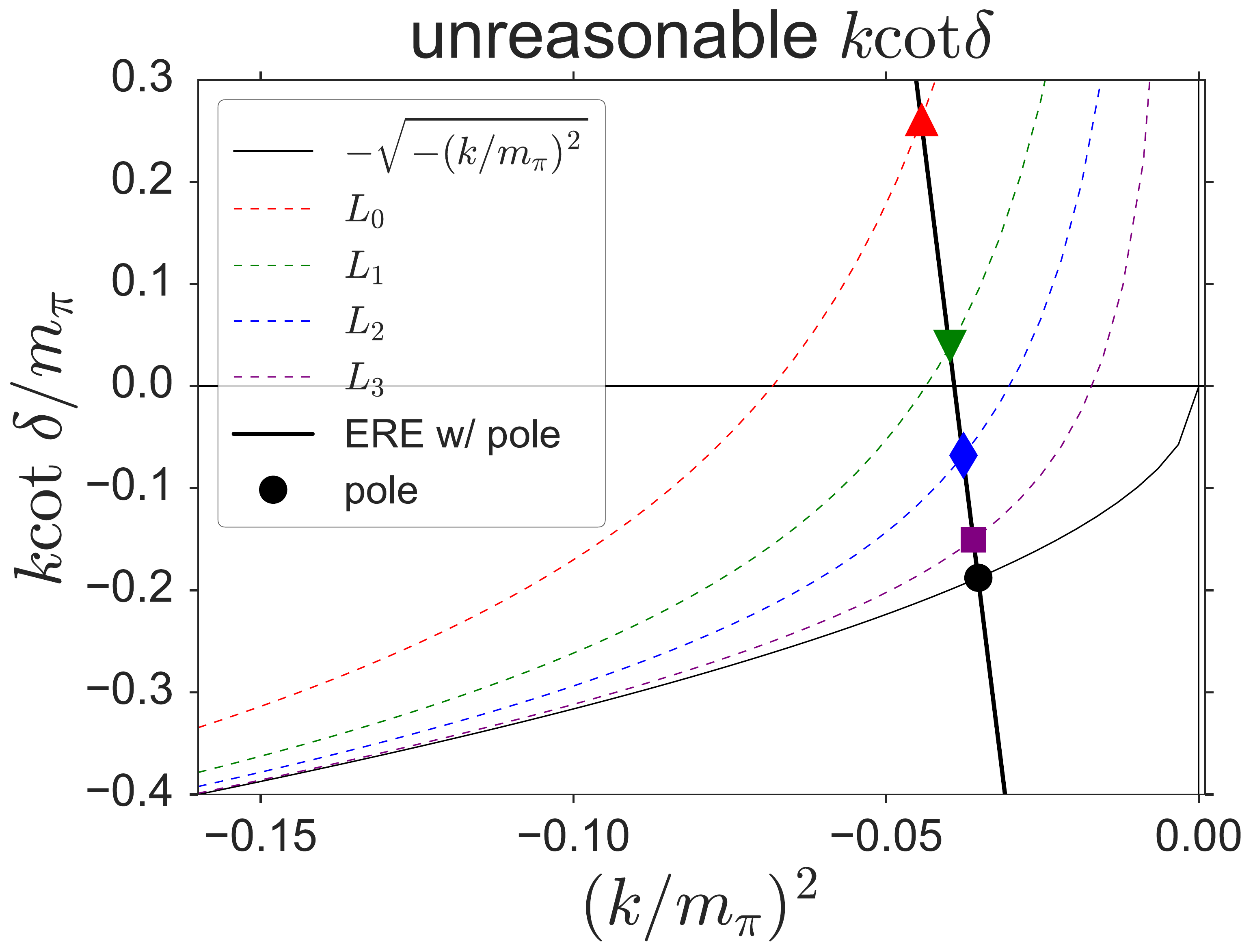}

 \caption{Illustration of the finite volume test.  (Left) If data of $k^2$ are correct, we observe a reasonable behavior of $k\cot\delta_0(k)$ on four volumes ($L_0< L_1<L_2 < L_3$), denoted by open or solid symbols depending on absence or existence of the bound state. (Right) If we observe the behavior such as ones given by solid symbols, on the other hand,
 we suspect that the plateaux which give this behavior are probably fake.
  }
 \label{fig:fake}
\end{figure}
Using this property of $k\cot\delta_0(k)$ at $k^2 <0$, we propose to use the formula to check whether the extracted energy shift $\Delta E$ and thus $k^2$ on the finite volumes are reliable or not. For example, $k^2$ extracted from the correct plateau show reasonable behaviors of $k\cot\delta_0(k)$, as illustrated in Fig.~\ref{fig:fake} (Left) by open or solid symbols, while the behavior of $k\cot\delta_0(k)$ like solid symbols in the right figure
brings us the strong suspicion that the plateaux which give these data are fake due to contaminations of excited states\cite{Iritani:2015dhu,Iritani:2016jie,Iritani:2016}.  Note however that a reasonable behavior itself can not guarantee its correctness. 
This test merely detects manifestly incorrect data of  $\Delta E$.

This formula also provides the most systematic way to extract the energy in the infinite volume limit for the bound state if exists.
One should fit data of  $k\cot\delta_0(k)$ by the ERE formula (\ref{eq:ERE}) with a few parameters, whose intersects with the bound state condition $k\cot\delta_0(k) = -\sqrt{-k^2}$ gives the binding energy in the infinite volume limit (the solid circle in the left figure).   None of previous studies except \cite{Berkowitz:2015eaa} took this strategy. Instead they used  the finite volume inspired formula\cite{Yamazaki:2011nd,Yamazaki:2012hi}, its approximation\cite{Beane:2011iw, Orginos:2015aya}, the constant fit\cite{Yamazaki:2015asa} or just taking the value at largest volume\cite{Beane:2012vq}.  

\section{Results}
We take data of $\Delta E$ from the original papers and convert  them to $k\cot\delta_0(k)/m_\pi$, which are plotted as a function of $(k/m_\pi)^2$. 

\subsection{YKU2011, YIKU2012 and YIKU2015}
\begin{figure}[tbh]	
\centering
  \includegraphics[width=0.46\textwidth]{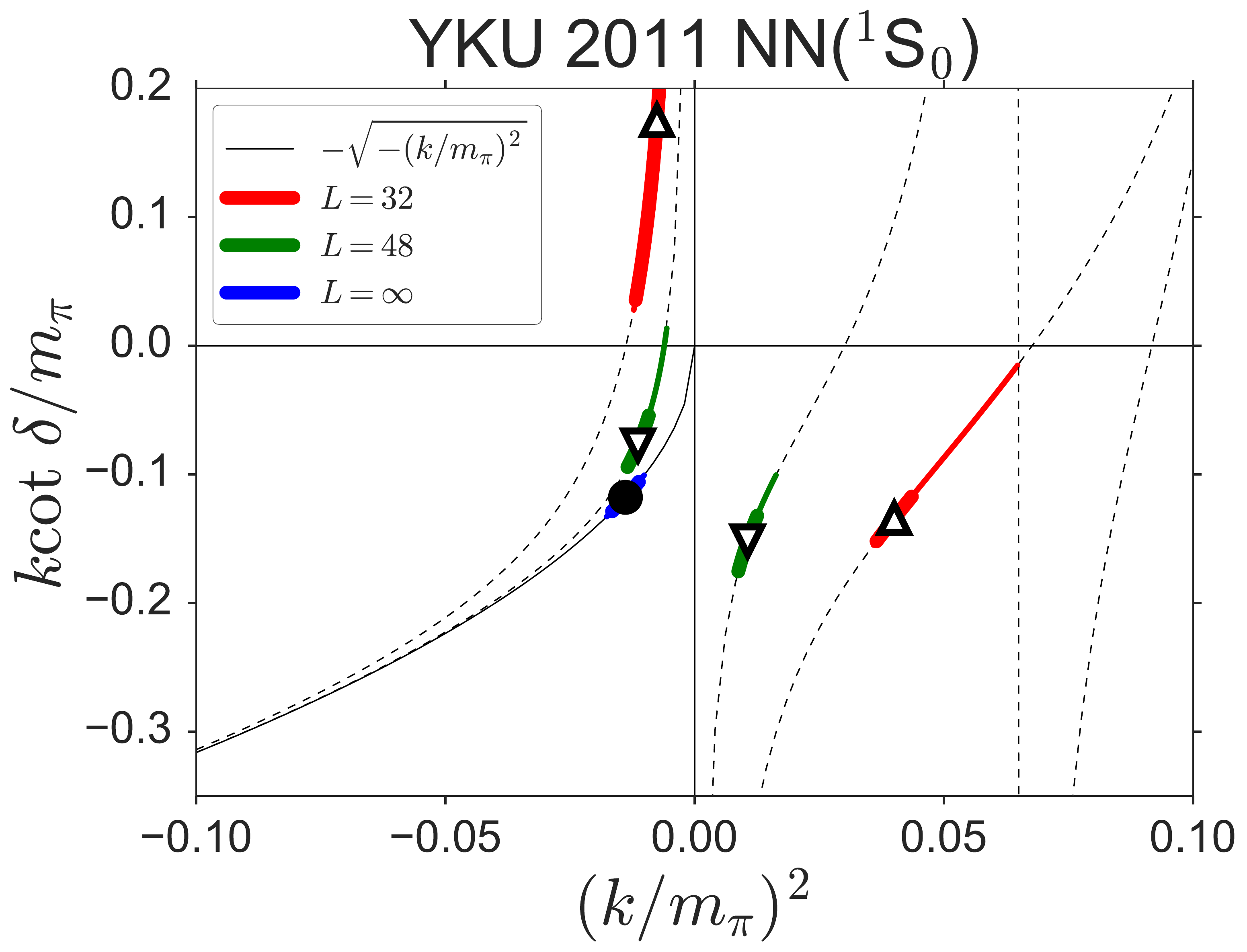}
   \includegraphics[width=0.46\textwidth]{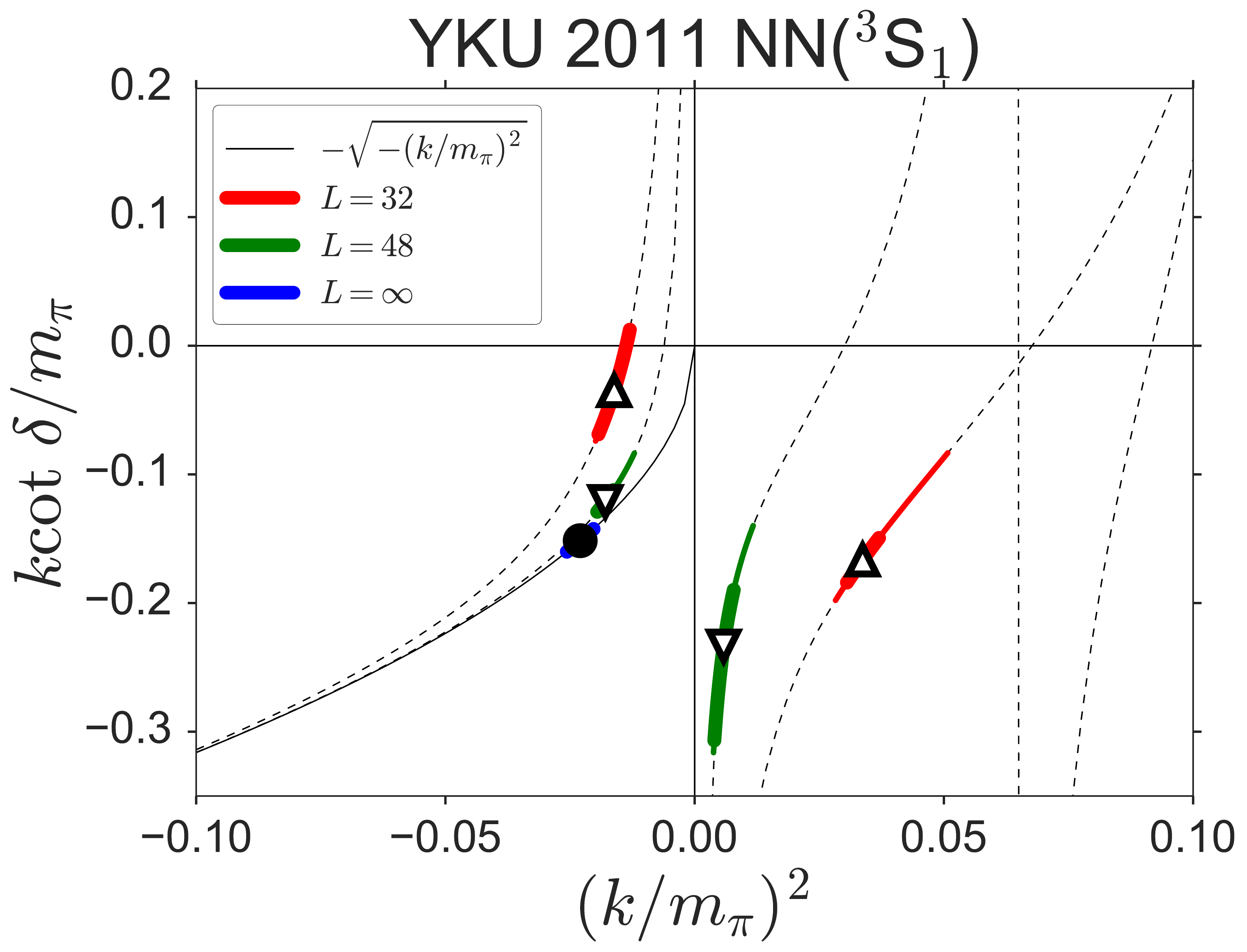}
     \includegraphics[width=0.46\textwidth]{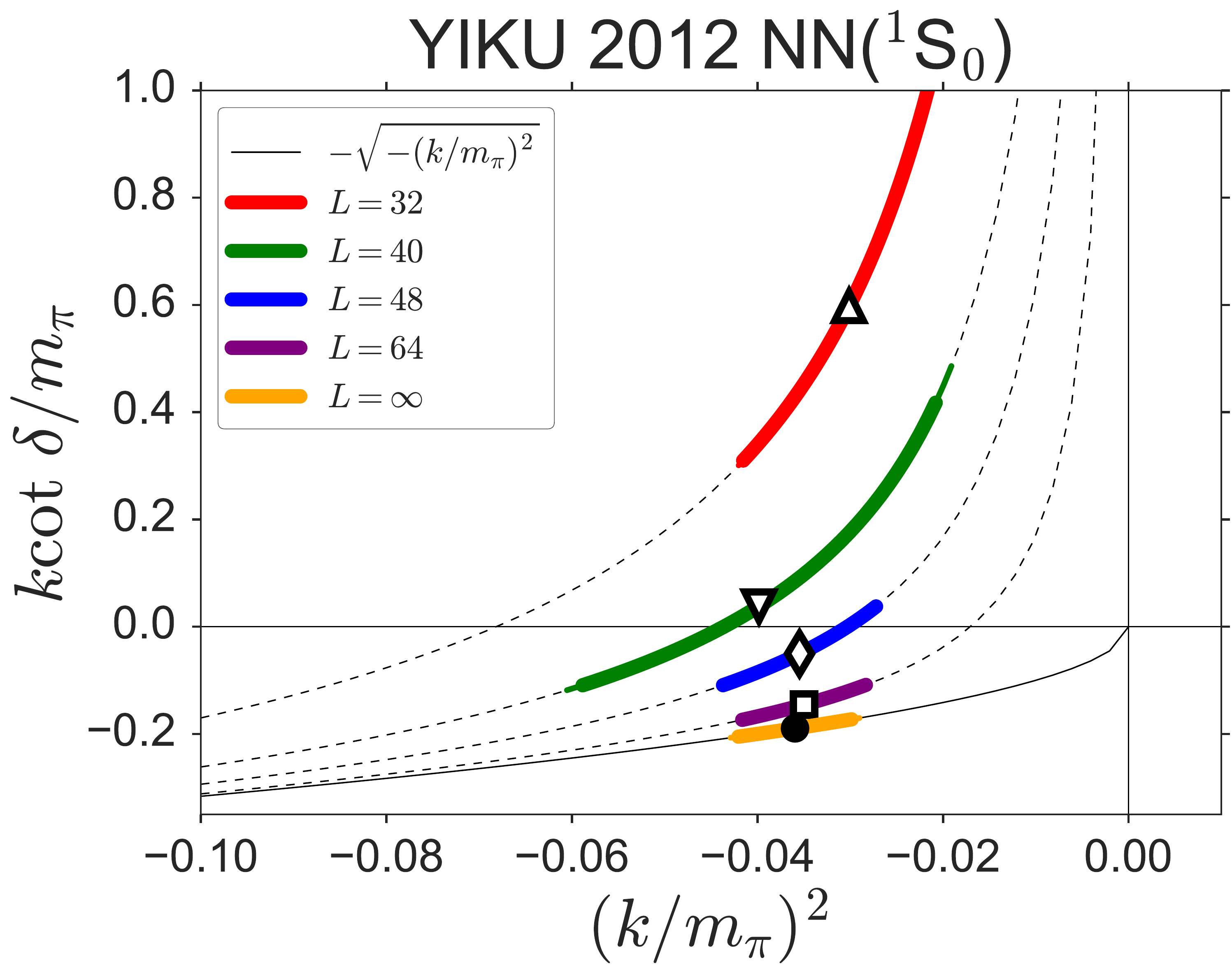}
   \includegraphics[width=0.46\textwidth]{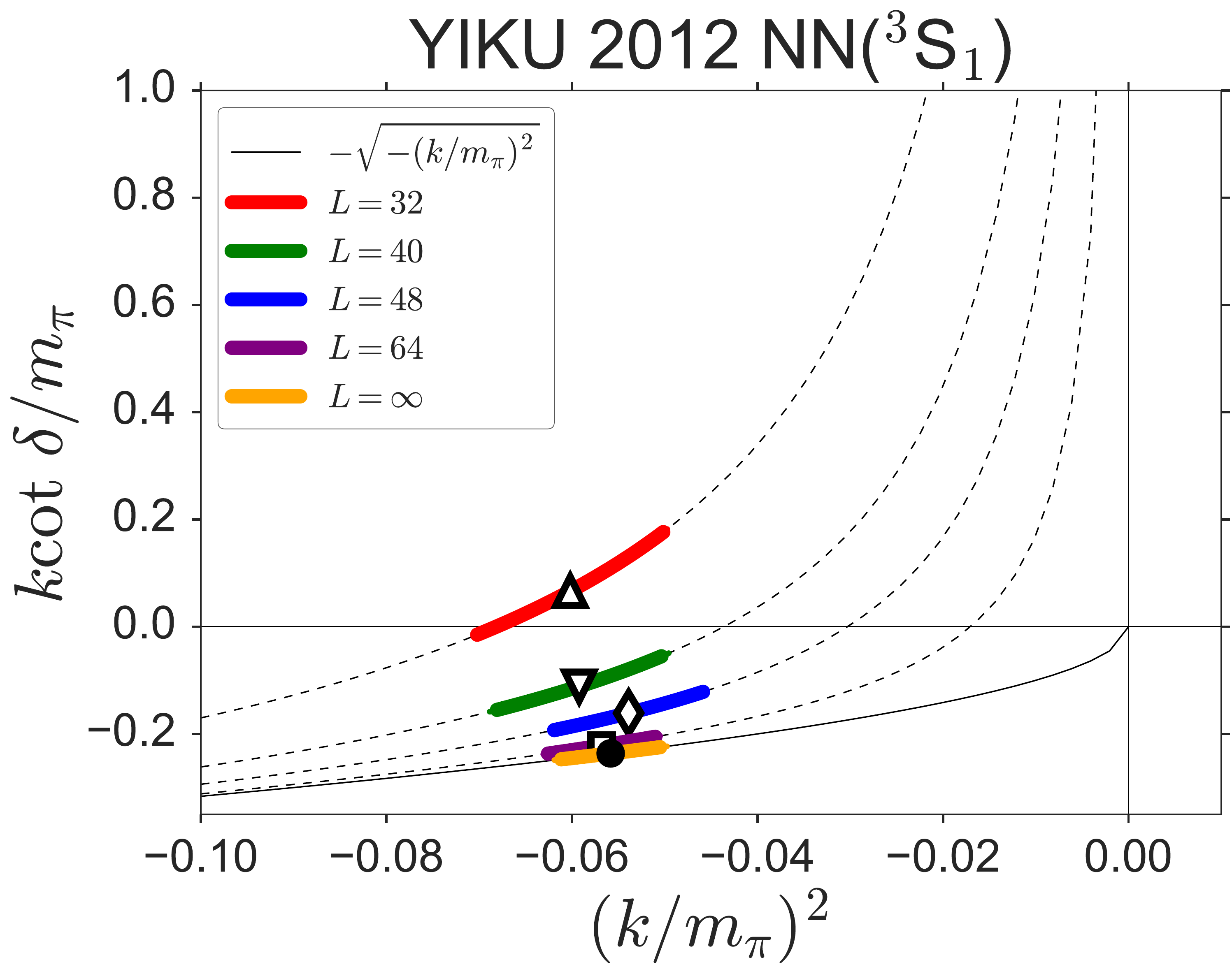}
     \includegraphics[width=0.46\textwidth]{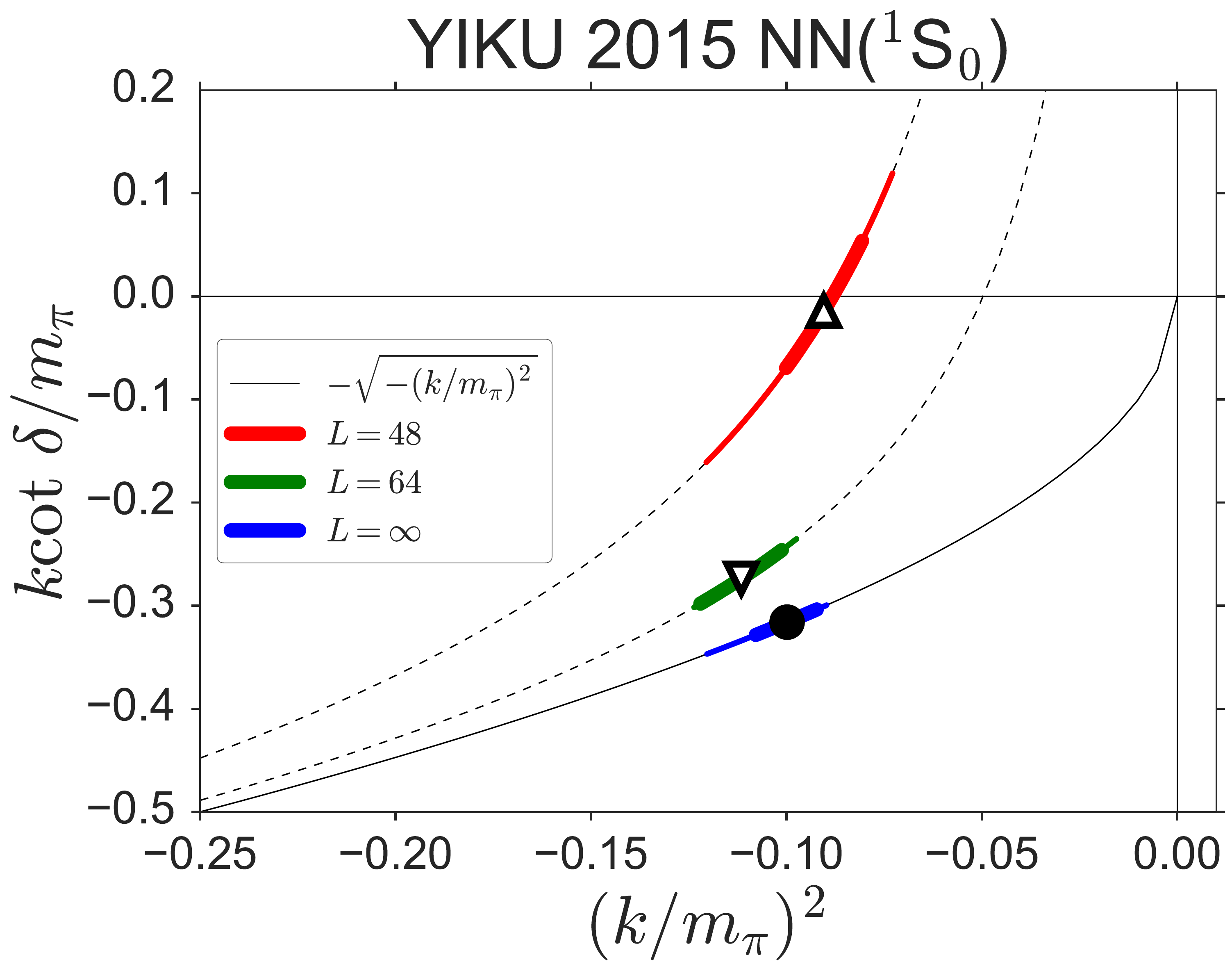}
   \includegraphics[width=0.46\textwidth]{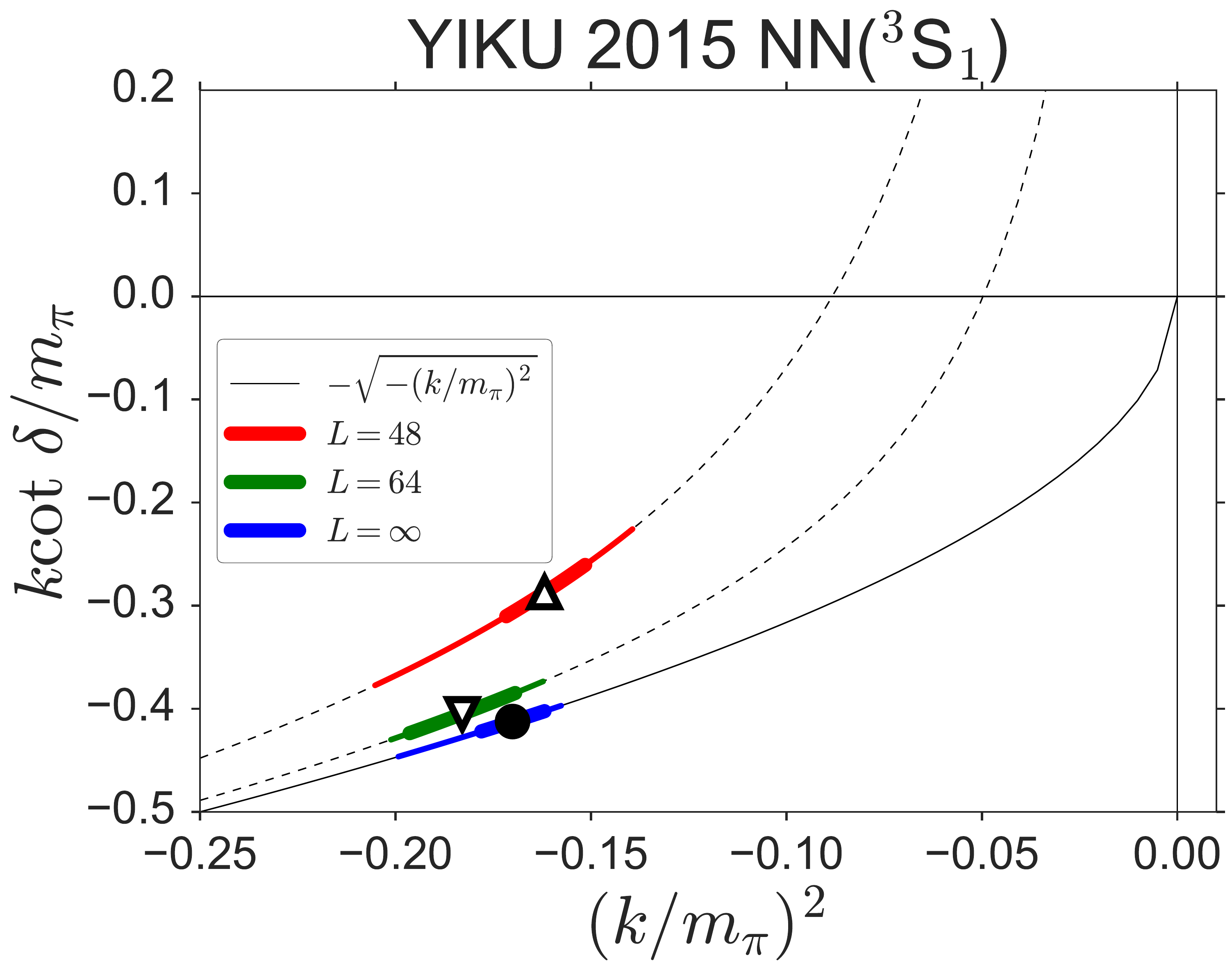}
 \caption{$k\cot\delta_0(k)/m_\pi$ as a function of $(k/m_\pi)^2$ for $NN$($^1$S$_0$) (Left) and
 $NN$($^3$S$_1$) (Right). Data from YKU2011 at $m_\pi=0.8$ GeV(Top), YIKU2012 at $m_\pi=0.51$ GeV (Middle) and YIKU2015 at $m_\pi=0.30$ GeV (Bottom).
 Dashed lines are  eq.~(\protect\ref{eq:kcot_delta}) while the black solid line represents the bound state condition that 
 $k\cot\delta_0(k)/m_\pi = -\sqrt{-(k/m_\pi)^2}$ in the infinite volume.
}
 \label{fig:kcot_YKU}
\end{figure}
We first consider YKU2011\cite{Yamazaki:2011nd}, YIKU2012\cite{Yamazaki:2012hi}  and YIKU2015\cite{Yamazaki:2015asa}, all of which claimed the existence of the $NN$ bound state for both $^1$S$_0$ and $^3$S$_1$ channels in quenched QCD at $m_\pi=0.8$ GeV\cite{Yamazaki:2011nd},  and 2+1 flavor QCD at $m_\pi=0.51$ GeV\cite{Yamazaki:2012hi} and $m_\pi=0.30$ GeV\cite{Yamazaki:2015asa}.

Fig.~\ref{fig:kcot_YKU} shows $k\cot\delta_0(k)/m_\pi$ as a function of $(k/m_\pi)^2$ for $NN({}^1{\rm S}_0)$ (Left) and
$NN({}^3{\rm S}_1)$ (Right), where data from both ground state and first excited state on $L/a=32,48$ are used for YKU2011 (Top) while data from only the ground state are employed for others (Middle and Bottom).
Black dashed lines in the figures represent the behavior of eq.~(\ref{eq:kcot_delta}) for each $L$, therefore uncertainties of $k\cot\delta_0(k)$ due to statistical (thick)  and systematic (thin) errors of $k^2$ follow these lines. The black solid line shows the bound state condition that $k\cot\delta_0(k)/m_\pi = -\sqrt{-(k/m_\pi)^2}$ in the infinite volume.

As can be seen from these figures, $k\cot\delta_0(k)/m_\pi$ shows a singular behavior for all 6 cases, 
which is very different from the behavior expected from the ERE in (\ref{eq:ERE}) at few lowest orders.
Since $(k/m_\pi)^2$ for these data are much smaller than 0.25, the expected convergence radius of the ERE, it is very unlikely that these singular behaviors in the figures are indeed true. In particular, behaviors of data at negative $k^2$ decrease almost vertically as the volume increases. These  behaviors are caused by the fact that extracted energy shift $\Delta E$ do not show significant volume dependences while the finite volume formula predicts dependences of $k\cot\delta_0(k)$ on these volumes.  The test here clearly indicates that the extracted energy shifts $\Delta E$ in these references are incorrect, due to the contamination from excited states nearby\cite{Iritani:2015dhu,Iritani:2016jie,Iritani:2016}.
We therefore conclude that $\Delta E$ in these references are not reliable, so that existences of $NN$ bound states at heavier pion mass  are not be established, contrary to their claims.
 
\subsection{NPL2012, NPL2013 and NPL2015}
\begin{figure}[tbh]	
\centering
  \includegraphics[width=0.46\textwidth]{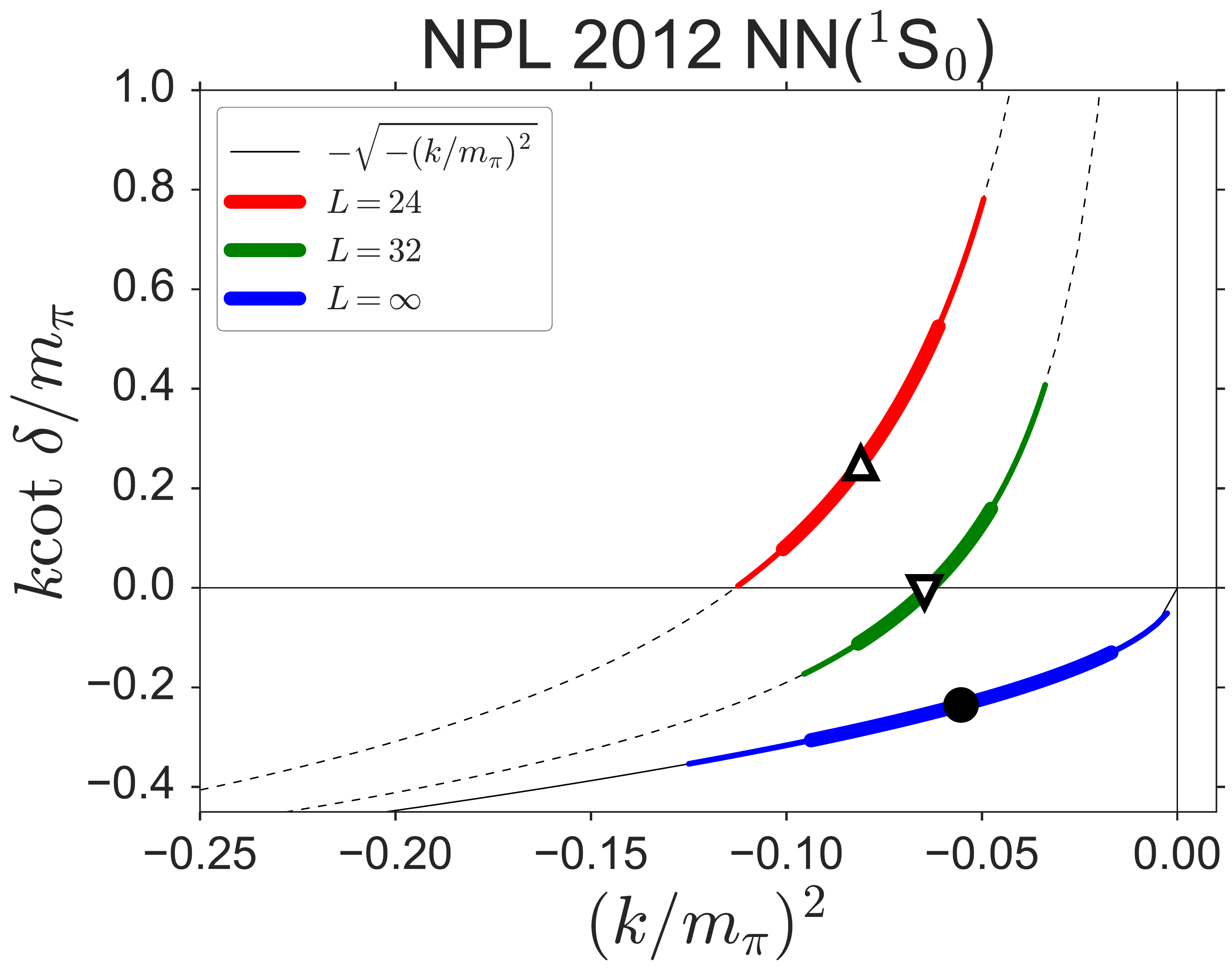}
   \includegraphics[width=0.46\textwidth]{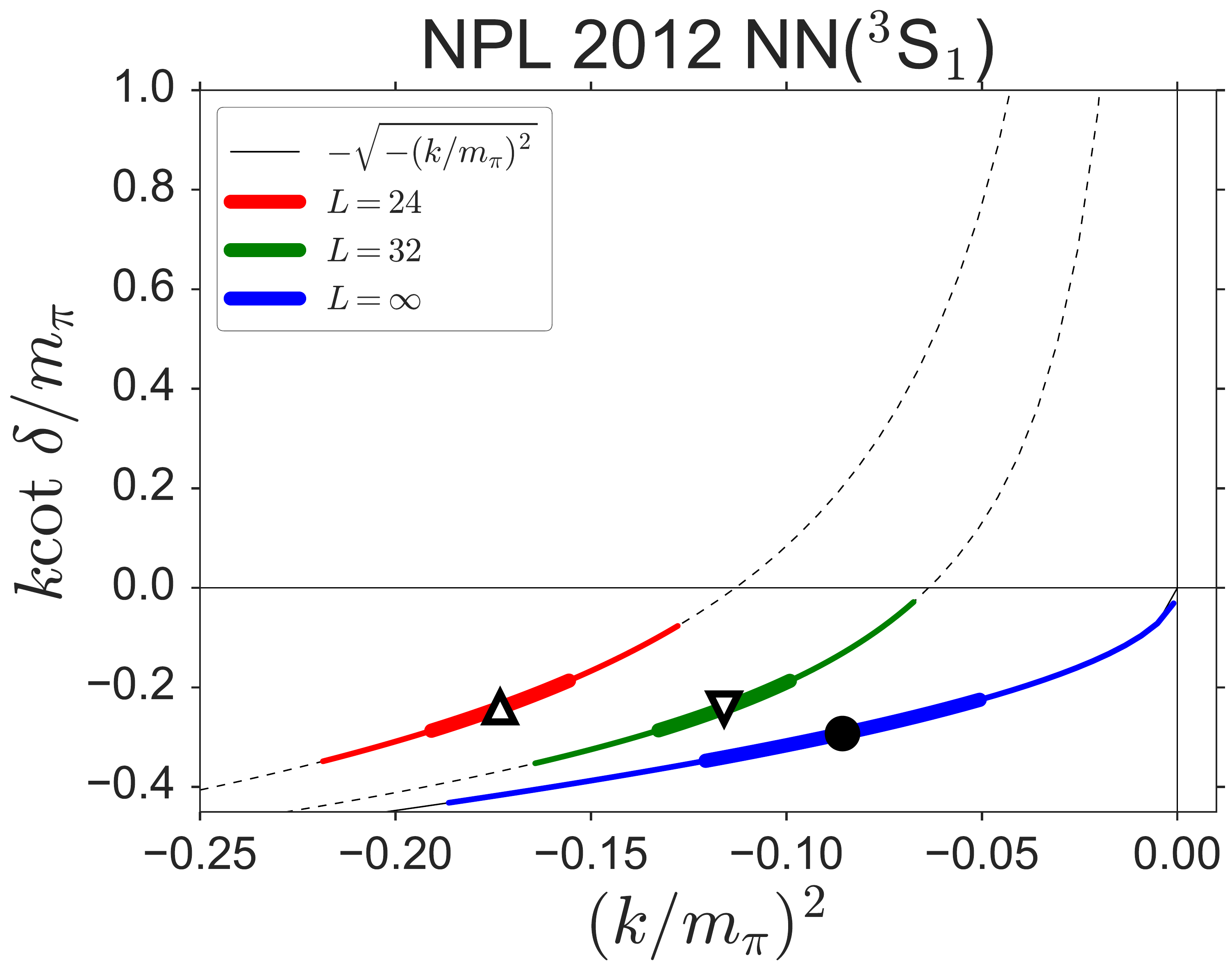}
     \includegraphics[width=0.46\textwidth]{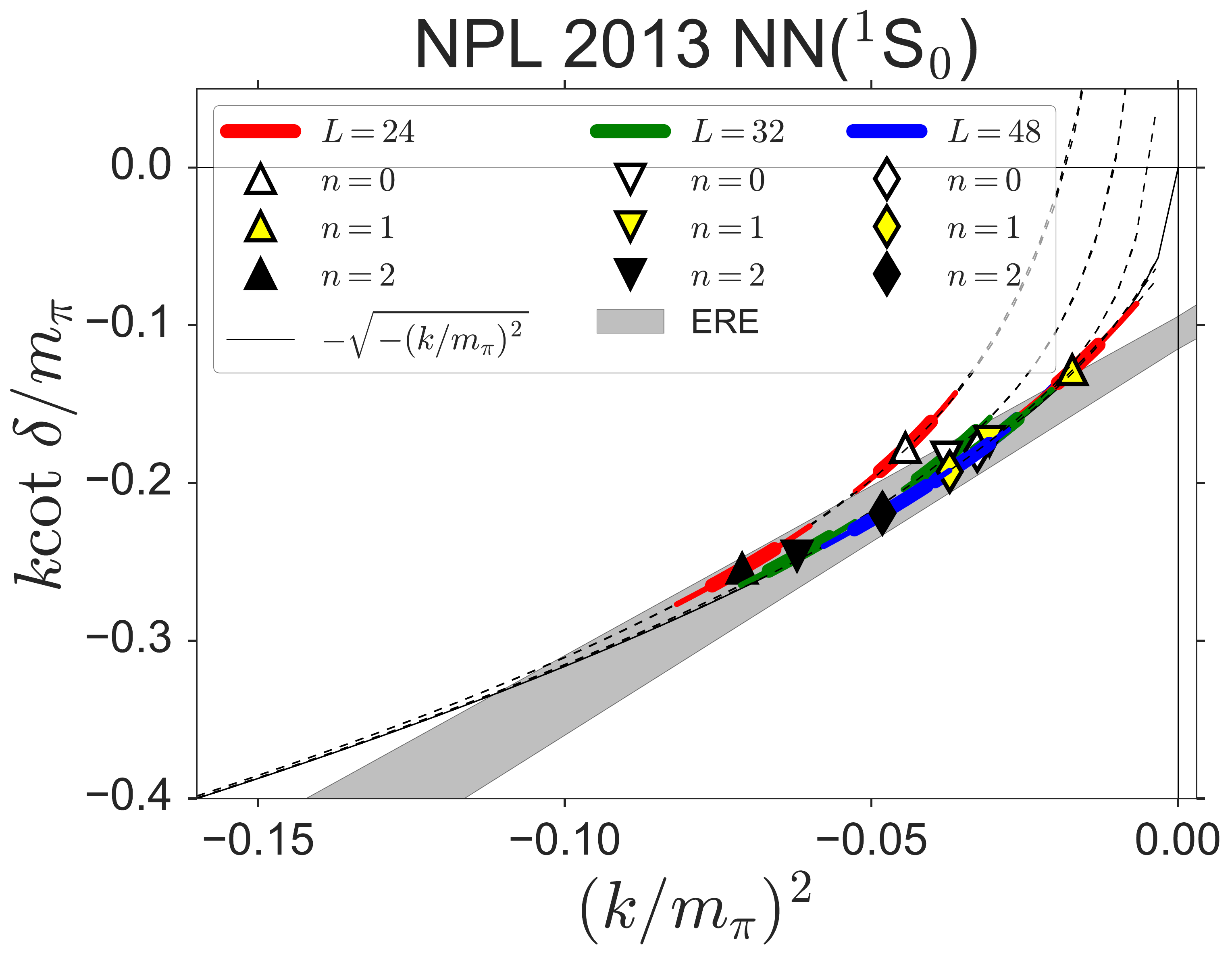}
   \includegraphics[width=0.46\textwidth]{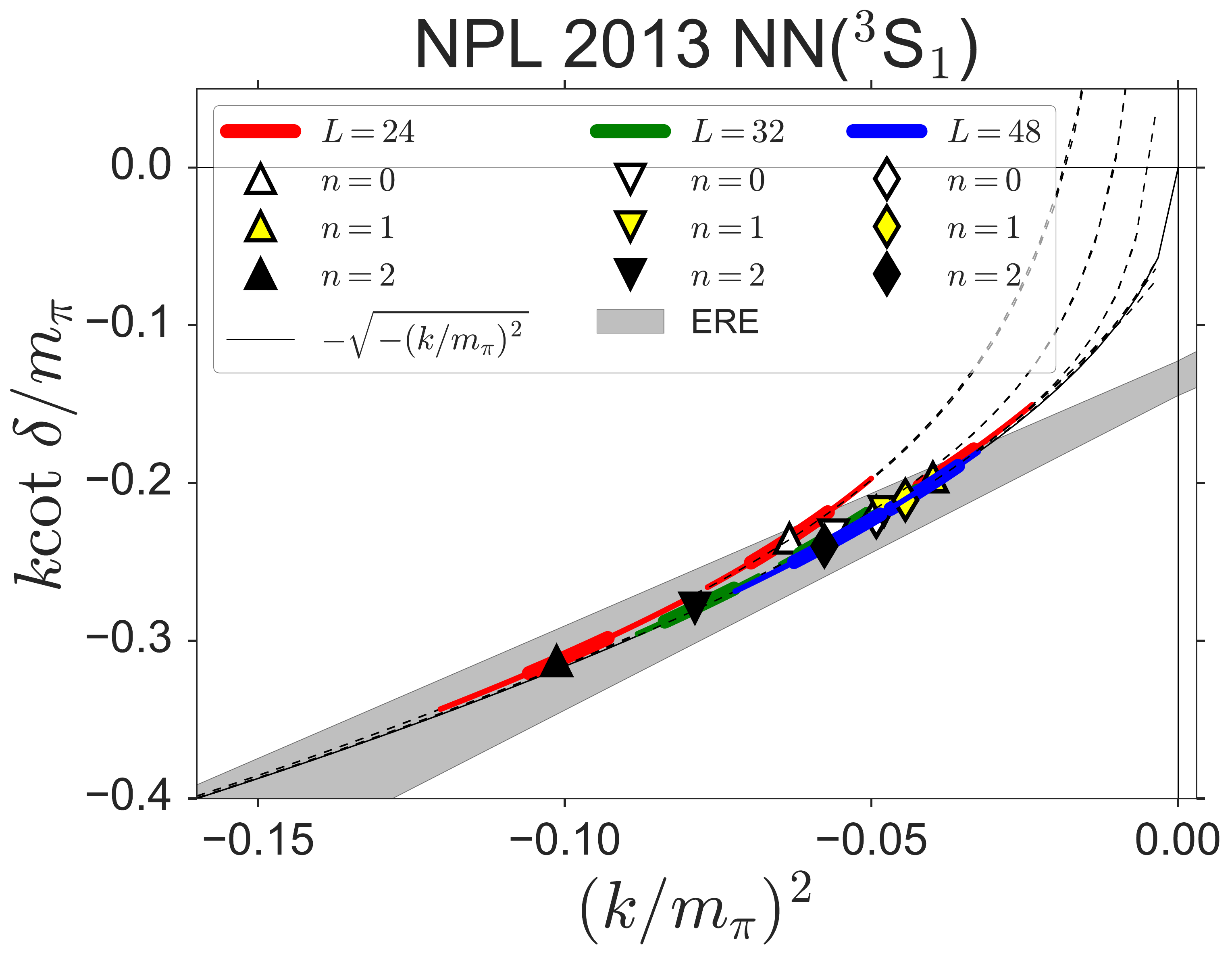}
  \includegraphics[width=0.46\textwidth]{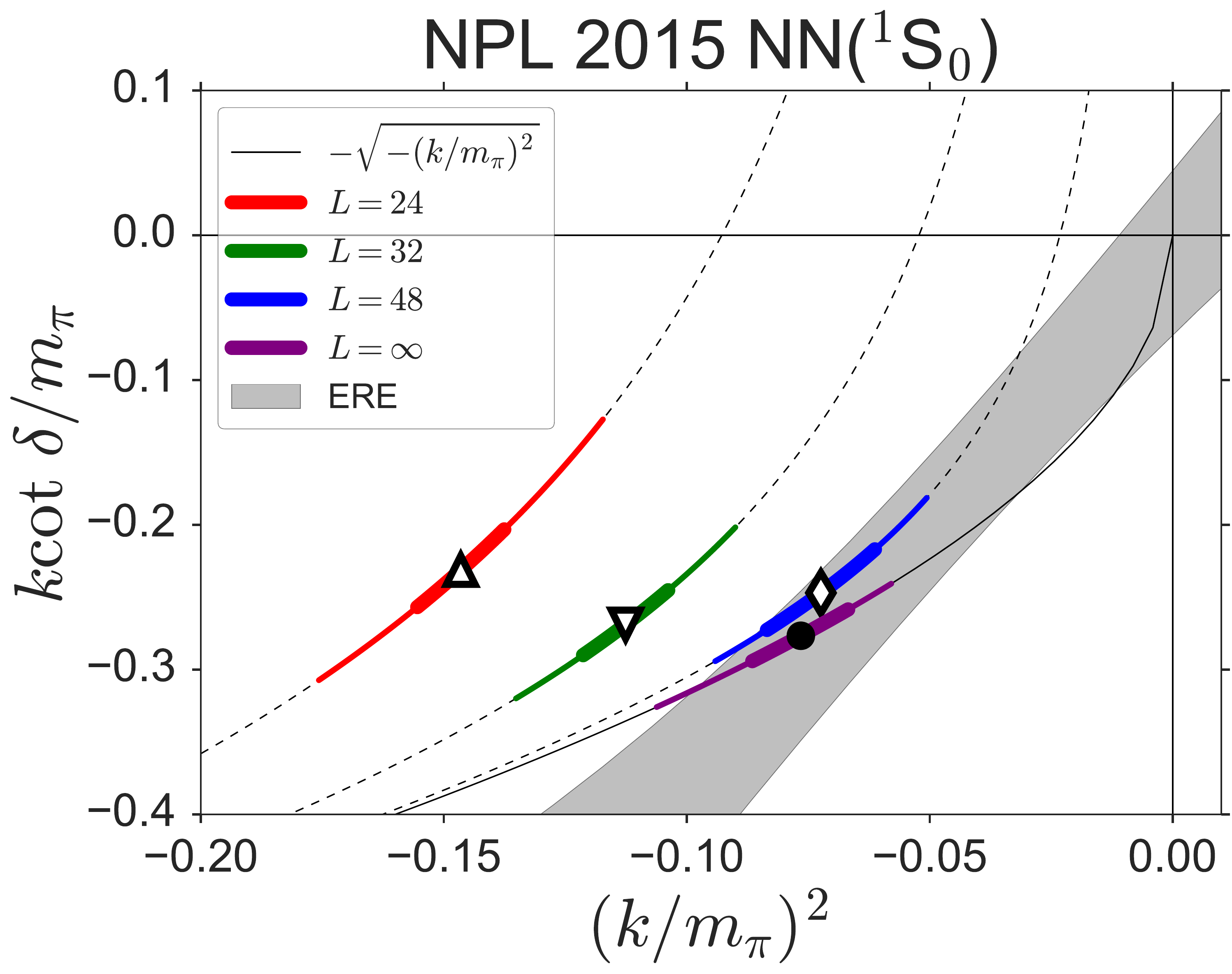}
   \includegraphics[width=0.46\textwidth]{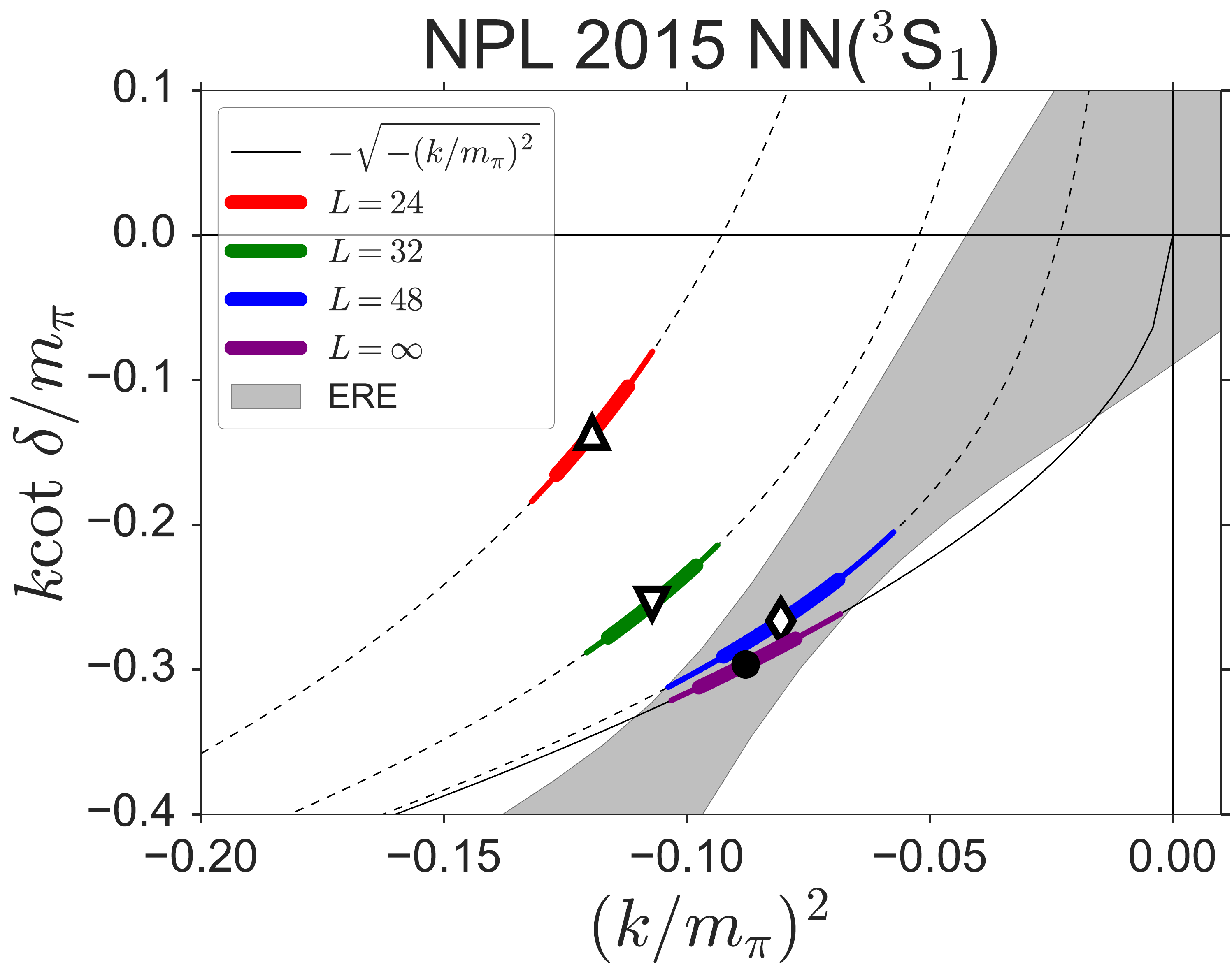}
 \caption{Same as Fig.~\protect\ref{fig:kcot_YKU} 
  for $NN$($^1$S$_0$) (Left) and $NN$($^3$S$_1$) (Right). 
  Data from NPL2012  at $m_\pi = 0.39$ GeV (Top), NPL2013 at $m_\pi=0.81$ GeV (Middle) and NPL2015 at $m_\pi=0.45$ GeV (Bottom). 
  Gray bands in the figures are their ERE in the literatures.
  }
 \label{fig:kcot_NPL2012-2015}
\end{figure}
Fig.~\ref{fig:kcot_NPL2012-2015} shows $k\cot\delta_0(k)/m_\pi$ for  $NN({}^1{\rm S}_0)$  (Left) and
$NN({}^3{\rm S}_1)$  (Right), from NPL2012\cite{Beane:2011iw} (Top), NPL2013\cite{Beane:2012vq} (Middle)
and NPL2015\cite{Orginos:2015aya} (Bottom). 

Although behaviors of data are less singular than those in the previous subsection, 
$k\cot\delta_0(k)/m_\pi$ at $k^2<0$ decreases as the volume increases. In particular, it decreases vertically for  
$NN({}^1{\rm S}_0)$  (Top-Left).
In addition, data at $k^2<0$ for $NN({}^1{\rm S}_0)$ (Bottom-Left) and  $NN({}^3{\rm S}_1)$  (Bottom-Right) suggest the negative effective range $r_0$, which seems inconsistent with the ERE fit to data at $k^2>0$ together with the binding energy in the infinite volume limit in the paper (gray band in the figure). Data at $k^2<0$ for $NN({}^1{\rm S}_0)$
 (Middle-Left) and $NN({}^3{\rm S}_1)$ (Middle-Right), obtained at boosted systems with the total momentum $P=n\times 2\pi /L$ ($n=1,2$) as well as  the center of mass system ($n=0$), look reasonable at first sight, but data at $n=0$ and $n=2$ on each $L$ disagree, albeit  the finite volume formula for $n=0,2$ almost overlap each other\cite{Rummukainen:1995vs}. 
Only data in figures (Top-Right) seems less problematic.
 
Although data are less singular, the finite volume test  brings a series suspicion that these data are affected by the fake plateau problem, as in the case of the previous subsection. Further investigation will be necessary to clear this suspicion.

\section{Conclusion}
In this report, we have proposed the finite volume test for the energy shift of two baryon systems and applied it to $NN$ data. We have found that data of YKU2011, YIKU2012 and 2015 show very singular ERE behaviors, suggesting that plateaux reported in these studies are probably  fake due to the contamination from excited states\cite{Iritani:2015dhu,Iritani:2016jie,Iritani:2016}. Results in this report, together with those in Ref.~\cite{Iritani:2015dhu,Iritani:2016jie,Iritani:2016}, conclude that existences of $NN$ bound states at heavier pion masses in these references are no more valid.

Although data from NPL2012, 2013 and 2015 are less singular, their reliabilities are still questionable.
Indeed, CalLat2015\cite{Berkowitz:2015eaa} reported an existence of two states at $k^2<0$ for $NN({}^3{\rm S}_1$ 
from two different source operators, using same gauge configurations of NPL2013.
We interpret this source dependence of states at $k^2<0$ as the manifestation of the fake plateaux problem\cite{Iritani:2015dhu,Iritani:2016jie,Iritani:2016}.

We  close this report by concluding that more careful studies are certainly necessary to establish the existence for the bound states of two or more baryon systems at heavier pion masses if any, contrary to optimistic views in previous  literatures. \\

This work is supported in part  by the Grant-in-Aid of the Japanese Ministry of Education, Sciences and Technology, Sports and Culture (MEXT) for Scientific Research (Nos. JP15K17667, JP16H03978) and by MEXT
and Joint Institute for Computational Fundamental Science (JICFuS)
as a priority issue ``Elucidation of the fundamental laws and evolution of the universe'' to be tackled by using Post K Computer. 
TI thanks Dr. Lorenzo Contessi for his suggestion, which triggered the study in this report.

\end{document}